\documentclass[12pt,twoside,a4paper]{article}
\usepackage[xdvi,dvips]{epsfig}
\usepackage{latexsym}
\usepackage{amsfonts}
\usepackage{amssymb}
\usepackage{amsmath}
\usepackage{enumerate}
\usepackage{exscale}
\usepackage{color}
\usepackage{graphicx}
\usepackage{ulem}

\newtheorem{theorem}{Theorem}[section]

\newtheorem{definition}{Definition}[section]

\title{{\bf On the Approximability of\\ Independent Set Problem on Power Law Graphs}}
\author{Mathias Hauptmann\thanks{Dept. of Computer Science, University of Bonn.
    e-mail:{ \tt hauptman@cs.uni-bonn.de}} \and
        Marek Karpinski\thanks{Dept. of Computer Science, University of Bonn and Hausdorff Center for Mathematics, University of Bonn.
        e-mail:{ \tt marek@cs.uni-bonn.de}}}
\date{}
\begin{document}
\maketitle
\begin{abstract}
We give the first nonconstant lower bounds for the approximability of the 
Independent Set Problem on the Power Law Graphs. These bounds are of the form 
$n^{\epsilon}$ in the case when the power law exponent satisfies $\beta <1$.
In the case when $\beta =1$, the lower bound is of the form $\log (n)^{\epsilon}$.
The embedding technique used in the proof
could also be of independent interest.
\end{abstract}
\nocite{aiello01}
\nocite{austrin11}
\nocite{chung02}
\nocite{khot}
\section{Introduction}

In this paper we prove new inapproximability results for the Maximum Independent Set (MIS) problem on Power Law Graphs. 

The Independent Set Problem on general graph instances is equivalent to the Max Clique Problem. 
In \cite{feige} it was shown that the Max Clique Problem is NP-hard to approximate within a factor $n^c$ for some constant $c>0$.
A randomized reduction from PCPs to Max Clique has been constructed in \cite{hastad}. This implies that if $\mbox{NP}\neq\mbox{ZPP}$, then 
Max Clique cannot be approximated within $n^{1-\epsilon}$ for any $\epsilon >0$. 
Khot \cite{khot} showed that under the Unique Games Conjecture (UGC), there exists 
some $\gamma >0$ such that Max Clique cannot be approximated within $n\slash 2^{(\log n)^{1-\gamma}}$. These results have been derandomized 
in \cite{zuckerman06}. This implies that unless $\mbox{P}=\mbox{NP}$, Max Clique cannot be approximated within $n^{1-\epsilon}$ for any $\epsilon >0$.
Furthermore, there exists some $\gamma >0$ such that, unless $\tilde{\mbox{P}}=\mbox{N}\tilde{\mbox{P}}$, 
no quasi-polynomial time algorithm approximates Max Clique within
$n\slash 2^{(\log n)^{1-\gamma}}$.

It has been discovered recently that many real-world large scale networks have a node degree distribution which follows a power law.
This has been observed for the graphs of the Internet \cite{faloutsos} and the World Wide Web (WWW) \cite{barabasi99}, 
peer-to-peer networks, gene regulatory networks and protein interaction networks \cite{bornholdt} and for social networks 
(\cite{redner}, \cite{kempe05}).

A random graph model for power law graphs has been introduced in \cite{aiello00a},\cite{aiello00b}.
Power law graphs have the property that their node degree distribution follows a power law, 
i.e. the number of nodes of degree $i$ is proportional to $i^{-\beta}$, for some fixed $\beta >0$.
This parameter $\beta$ is called the power law exponent. 

It has been observed experimentally that some optimization problems tend to be much easier to solve on power law graphs than on general graph instances
\cite{eubank}.
This rises the question on whether one can show differences in terms of approximability or approximation hardness of several optimization problems 
between general graph instances and power law graphs \cite{shen12}. 
Ferrante et al. \cite{ferrante} showed the NP-hardness of Max Clique in power law graphs for $\beta\geq 1$ and of MIS in power law graphs for all $\beta >0$. 
Shen et al. \cite{shen12} proved the APX-hardness of Max Clique and Maximum Independent Set on power law graphs, also for $\beta >1$. 
Their result is based on an efficient embedding of bounded degree graphs into power law graphs. 

We consider the Maximum Independent Set (MIS) problem in Power Law Graphs (PLGs) for $\beta\leq 1$. 
For $\beta\in (0,1)$ we show that the MIS problem in $(\alpha,\beta )$-PLGs is hard to approximate within $n^{\epsilon}$, for some $\epsilon >0$ being constant.
For $\beta =1$, we give a lower bound of $\Delta^{\epsilon_2}$ for some $\epsilon_2>0$, where $\Delta=e^{\alpha}$ is the maximum degree of a power law graph.

The paper is organized as follows. In Section 2 we give the formal definition of a power law graph according to \cite{aiello00a}.
Section 3 provides an important tool which we will use in our constructions, namely a method how to complete constructively fragments of a given power law
node degree distribution such that the maximum independent set size of the resulting graph is small.
In Section 4 we will use this auxiliary construction to obtain our hardness result for the case when $\beta <1$. The case $\beta =1$
is considered in Section 5. In Section 6 we give a summary and a number of open questions.

\section{Power Law Graphs}
In this section we start with giving the formal definition of the $(\alpha,\beta )$ Power-Law Graphs (cf. \cite{aiello00a},\cite{aiello00b}).
We introduce also the notion of a node degree interval, which will be of particular importance within our constructions.
Informally, a {\sl node degree interval} (or just {\sl interval} for short) is a subset of nodes in a graph whose node degrees 
are all within a given interval $[a,b]$. In the subsequent sections we will be concerned with the construction of embeddings of a given graph
$G$ into some node degree interval of a power law graph.  
In this section we provide estimates for 
sizes and volumes (i.e. sums of node degrees) of intervals in $(\alpha,\beta )$ Power-Law Graphs (PLGs).
\begin{definition}\cite{aiello01}
An undirected multigraph $G=(V,E)$ with self loops is called an $(\alpha,\beta)$ Power-Law Graph if the following conditions hold:
\begin{itemize}
\item The maximum degree is $\Delta=\lfloor e^{\alpha\slash\beta}\rfloor$.
\item For $i=1,\ldots , \Delta$, the number $y_i$ of nodes of degree $i$ in $G$ satisfies
      \[y_i\: =\: \left\lfloor \frac{e^{\alpha}}{i^{\beta}}\right\rfloor\] 
\end{itemize}
\end{definition}
The following estimates for the number $n$ of vertices of an $(\alpha,\beta)$ Power-Law Graph are well known (cf. \cite{aiello01}):
\[n\:\approx\: \left\{\begin{array}{l@{\quad}l}
 \frac{e^{\alpha\slash\beta}}{1-\beta} & \mbox{for $0<\beta <1$,}\\
 \alpha\cdot e^{\alpha} & \mbox{for $\beta =1$,}\\
 \zeta (\beta )\cdot e^{\alpha} & \mbox{for $\beta >1$.}
\end{array}\right.\quad 
m\:\approx\: \left\{\begin{array}{l@{\quad}l}
 \frac{1}{2}\frac{e^{2\alpha\slash\beta}}{2-\beta} & \mbox{for $0<\beta <2$,}\\
 \frac{1}{4}\alpha e^{\alpha} & \mbox{for $\beta =2$,}\\
 \frac{1}{2}\zeta (\beta -1)e^{\alpha} & \mbox{for $\beta >2$.}
\end{array}\right.\]
Here $\zeta (\beta)=\sum_{i=1}^{\infty}i^{-\beta}$ is the {\sl Riemann Zeta Function}.

Given an $(\alpha,\beta)$ Power-Law Graph $G=(V,E)$ with $n$ vertices and maximum degree $\Delta$ and two integers $1\leq a\leq b\leq\Delta$, 
an {\sl interval} $[a,b]$ is defined as the subset of $V$
\begin{equation}
[a,b]\:\: =\:\: \{v\in V|a\leq \mbox{deg}_G(v)\leq b\}
\end{equation}
If $U\subseteq V$ is a subset of vertices, the {\sl volume} $\mbox{vol}(U)$ of $U$ is defined as the sum of node degrees of nodes in $U$.
We have the following estimates for sizes and volumes of node intervals in $(\alpha,\beta)$-PLGs (cf. \cite{ghk_domset}).
\begin{theorem}\cite{ghk_domset}
For $0<\beta <1$, we have
\[\begin{array}{l@{\:}c@{\:}l}
|[x\Delta,y\Delta ]| & \in & \left [\frac{\Delta}{1-\beta}(y^{1-\beta}-x^{1-\beta})
                                    -\left (\frac{1}{x^{\beta}}-\frac{1}{y^{\beta}}\right ),\:
                                    \frac{\Delta}{1-\beta}(y^{1-\beta}-x^{1-\beta})\right ]\\
|[x\Delta,\Delta]| & \in & \left [\frac{\Delta}{1-\beta}(1-x^{1-\beta})
                           -\left (\frac{1}{x^{\beta}}-1\right ),\:
                           \frac{\Delta}{1-\beta}(1-x^{1-\beta})\right ]\\
\mbox{vol}([x\Delta,\Delta]) & \geq & \Delta^2\left (\frac{1-x^{2-\beta}}{2-\beta}-\frac{1}{2}+\frac{x^2}{2}\right )
                                      -\Delta\left (1-x^{1-\beta}-\frac{1}{2}+\frac{x}{2}\right )
\end{array}\]
For $\beta=1$, we have
\[\begin{array}{l@{\:}c@{\:}l}
|[x\Delta,y\Delta ]| & \in & \left [e^{\alpha}\cdot\left (\ln\left (\frac{1}{x}\right )-\ln\left (\frac{1}{y}\right )\right )-(y-x)e^{\alpha},\: 
                                    e^{\alpha}\cdot\left (\ln\left (\frac{1}{x}\right )-\ln\left (\frac{1}{y}\right )\right )\right ]\\
\mbox{vol}([x\Delta,y\Delta ]) & \in & \left [e^{\alpha}(y-x)\Delta -\left (\frac{y\Delta (y\Delta +1)}{2}-\frac{x\Delta (x\Delta +1)}{2}\right ),
                                              e^{\alpha}(y-x)\Delta\right ]
\end{array}\]
\end{theorem}
These estimates will be used in subsequent sections where we construct efficient reductions from the general Maximum Independent Set (MIS) problem 
to MIS in $(\alpha,\beta )$-PLGs for $\beta\leq 1$.
In particular we will be concerned with the embedding of the node degree distribution of a given graph $G$ into an interval of the form $[x\Delta,\Delta ]$ of a 
power law node degree distribution with parameters $\alpha$ and $\beta$.
\section{An Auxiliary Graph Construction}
We start with an auxiliary problem.

Our hardness results for the MIS problem in Power Law Graphs rely on the construction and analysis of embeddings
of a given class of graphs $G$, namely a class of instances on which the MIS problem is known to be hard to approximate
into $(\alpha, \beta)$-PLG. Such an instance $G$ will be mapped to a node degree interval of the form $[x\Delta,\Delta ]$, where the 
parameter $\alpha$ (and hence $\Delta =e^{\alpha\slash\beta}$) has been chosen appropriately. Then the task will be to 
construct the remaining part of the power law graph, which consists of a subgraph corresponding to the interval $[1,x\Delta -1]$
and the rest of the interval $[x\Delta ,\Delta]$.
This leads us to the following construction problem.
We are given a sequence of node degrees $d=(d_1,\ldots , d_n)$. We want to construct a multigraph $G_d$ with $n$ vertices such that $d$ is the degree
sequence of $G_d$ and furthermore, the size $IS(G_d)$ of a maximum independent set in $G_d$ is {\sl as small as possible}.
In particular we have to consider the case when the degree sequence $d$ corrsponds to an {\sl interval} $[a,b]$ of a power law distribution
with $1\leq a\leq b\leq\Delta=\lfloor e^{\alpha\slash\beta}\rfloor$.
This means that for each $a\leq i\leq b$, the sequence $d$ contains $\lfloor\frac{e^{\alpha}}{i^{\beta}}\rfloor$ entries all equal to $i$. 
We may assume that the sequence $d$ is always {\sl sorted}, i.e.
the entries in $d$ are in increasing order $d_1\leq\ldots \leq d_n$.

The idea of the construction is as follows. We cover the vertices of the auxiliary graph $G_d$ by a small number $n_c$ of cliques, 
in increasing order by their node degrees. Then $\lceil n\slash n_c\rceil$ is an
upper bound for $IS(G_d)$. Let us now describe this construction in more detail.

We consider the following situation. We are given a node degree interval 
$[a,b]$ with $1\leq a<b\leq\Delta=\left\lfloor e^{\alpha\slash\beta}\right\rfloor$.
We want to construct a graph $G_{a,b}$ such that the degree distribution of this graph is precisely equal to
the part of the power law node degree distribution corresponding to this interval.
Moreover, we want to achieve that, informally, the size $\mbox{IS}(G_{ab}$ of a maximum independent set in the graph $G_{a,b}$ is sufficiently small.

Let us now describe our construction in detail.
From the node degree interval $[a,b]$ we first construct the associated degree sequence 
$(d_1,\ldots , d_m)$ with $m=\sum_{j=a}^b\left\lfloor\frac{e^{\alpha}}{j^{\beta}}\right\rfloor$.
The set of vertices is
\[V_{a,b}=\{v_1,\ldots , v_m\}\quad\mbox{for $m=\sum_{j=a}^b\left\lfloor\frac{e^{\alpha}}{j^{\beta}}\right\rfloor$}\]
The degree sequence $(d_1,\ldots , d_m)$  has the following form:
\[\begin{array}{l}
d_1=\ldots = d_{\lfloor e^{\alpha}\slash a^{\beta}\rfloor}=a,\\
d_{\lfloor e^{\alpha}\slash a^{\beta}\rfloor +1} =\ldots 
= d_{\lfloor e^{\alpha}\slash a^{\beta}\rfloor + e^{\alpha}\slash (a+1)^{\beta}\rfloor}=a+1
\end{array}\]
and in general, if $j$ is the last index of a node of degree $i$, then the nodes
with indices $j+1,\ldots , j+\lfloor e^{\alpha}\slash (i+1)^{\beta}\rfloor$ are of degree $i+1$.
 
We generate the set of edges as follows:
\begin{itemize}
\item We take the first $d_1+1$ nodes $v_1,\ldots , v_{d_1+1}$ and connect them by a clique. 
\item We compute residual degrees accordingly.
\item We take the next $d_{d_1+2}$ nodes, connect them by a clique, keep track of residual degrees and iterate.
\end{itemize}
So in each iteration $i$, we construct a clique of size $d_{p(i)}+1$ on the set of nodes $\{v_{p(i)},\ldots , v_{p(i)+d_{p(i)}}\}$.
The function $p(\cdot )$ satisfies:
\begin{itemize}
\item $p(1) =1$ 
\item $p(i+1)=p(i)+d_{p(i)}+1$
\end{itemize}
We give an upper bound on the size of an independent set in this graph $G_{a,b}$.
If we would cover each set $V_i=\{v\in V(G_{a,b})|\mbox{deg}(v)=i\}$ separately, the number of cliques needed for $V_i$ would be
bounded by
\begin{equation}
\left\lceil \frac{\left\lfloor \frac{e^{\alpha}}{i^{\beta}}\right\rfloor}{i} \right\rceil \:\:\leq\:\:
\frac{\left\lfloor \frac{e^{\alpha}}{i^{\beta}}\right\rfloor+i}{i}\: \leq\: \frac{\frac{e^{\alpha}}{i^{\beta}}+i}{i}
\:\: =\:\: \frac{e^{\alpha}+i^{\beta +1}}{i^{\beta +1}}
\end{equation}
Thus the size of a maximum independent set in this graph is bounded by
\begin{eqnarray*}
\sum_{i=a}^{b}\frac{e^{\alpha}+i^{\beta +1}}{i^{\beta +1}} &  \leq &
\int_{a}^{b+1}\frac{e^{\alpha}}{i^{\beta +1}}dx\: +\:
\frac{e^{\alpha}}{a^{\beta +1}}-\frac{e^{\alpha}}{(b+1)^{\beta +1}}+b+1-a\\
 & = &  \left [\frac{e^{\alpha}}{-\beta\cdot x^{\beta}}\right ]_{a}^{b+1} \: +\:
        \frac{e^{\alpha}}{a^{\beta +1}}-\frac{e^{\alpha}}{(b+1)^{\beta +1}}+b+1-a\\
 & = &  \frac{e^{\alpha}}{\beta}\cdot\left (\frac{1}{a^{\beta}}-\frac{1}{(b+1)^{\beta}}\right )\: +\:
        \frac{e^{\alpha}}{a^{\beta +1}}-\frac{e^{\alpha}}{(b+1)^{\beta +1}}+b+1-a
\end{eqnarray*}
The number of nodes of the graph $G_{a,b}$ is $\sum_{i=a}^b\left\lfloor \frac{e^{\alpha}}{i^{\beta}}\right\rfloor$, which is contained in 
the interval
\begin{eqnarray*} & &  \left [ \frac{\Delta}{1-\beta}\left (\left (\frac{b}{\Delta}\right )^{1-\beta}-\left (\frac{a}{\Delta}\right )^{1-\beta}\right ) 
        -\left (\frac{\Delta^{\beta}}{a^{\beta}}-\frac{\Delta^{\beta}}{b^{\beta}}\right ),\:
        \frac{\Delta}{1-\beta}\left (\left (\frac{b}{\Delta}\right )^{1-\beta}-\left (\frac{a}{\Delta}\right )^{1-\beta}\right ) \right ]\\
& & = \: \left [\frac{e^{\alpha}}{1-\beta}\left (b^{1-\beta}-a^{1-\beta}-\frac{1}{a^{\beta}}+\frac{1}{b^{\beta}} \right ),
                \frac{e^{\alpha}}{1-\beta}\left (b^{1-\beta}-a^{1-\beta}\right ) \right ]
\end{eqnarray*} 
These estimates will be used in Section 4 and Section 5 in the analysis of our embeddings of graphs into $(\alpha, \beta)$ PLGs.

\section{An Embedding of Graphs into PLGs for $\beta<1$}
In this section we show that for each $\beta <1$, the MIS problem on $(\alpha,\beta )$-PLGs is NP-hard to approximate within 
$n^{\epsilon}$ for some constant $\epsilon\in (0,1)$ which only depends on $\beta$.  
This result is based on the construction of an efficiently computable embedding of arbitrary graphs into $(\alpha,\beta )$-PLGs for $\beta <1$. 
The global structure of this embedding is as follows. Given a graph $G$ which we want to embed, we map the vertices of $G$ to a node degree interval
of the form $[x\Delta, \Delta ]$ of some power law distribution with parameters $\alpha,\beta$, where $\Delta=\lfloor e^{\alpha\slash\beta }\rfloor$.
Then we make use of our auxiliary graph construction from Section 3 in order to construct the remaining parts of the power law graph $G_{\alpha,\beta}$.
By a careful choice of the parameters $x$ and $\alpha$ of this construction, we will be able to bound the size of a maximum independent set in the
residual graph $G_{\alpha,\beta }\setminus G$ such that the approximation hardness carries over from general graph instances to  $(\alpha,\beta )$-PLGs.

Given a graph $G$ with $m$ nodes, we start by replacing $G$ by the graph $G'$ which contains 
for each node $v_i$ in $G$ a clique of size $2$ consisting of nodes $v_{i,1},v_{i,2}$. Now for all $1\leq i<j\leq m$, nodes $v_{i,k}$ and $v_{j,l}$ ($k,l\in\{1,2\}$) are connected
by an edge iff $G$ contains an edge between $v_i$ and $v_j$. The number of nodes of $G'$ is $n=2m$, and we have $IS(G')=IS(G)$. Furthermore $G'$ contains a perfect matching,
consisting of edges $\{v_{i,1},v_{i,2}\}\: (1\leq i\leq m)$. This will enable us to replace these edges by multi-edges in order to fit the graph into
some given part of a power law node distribution.
We will now construct an embedding of the graph $G'$ into an $(\alpha,\beta )$-PLG.
In particular, the nodes of $G'$ will be mapped to nodes in the node interval $[x\Delta, \Delta]$, where $x$ is a parameter of the construction.
It turns out that we can choose $\alpha$ and $x$ in such a way that $n\geq \frac{1}{2}\cdot |[x\Delta, \Delta]|$.

\begin{figure}[htb]
\begin{center}
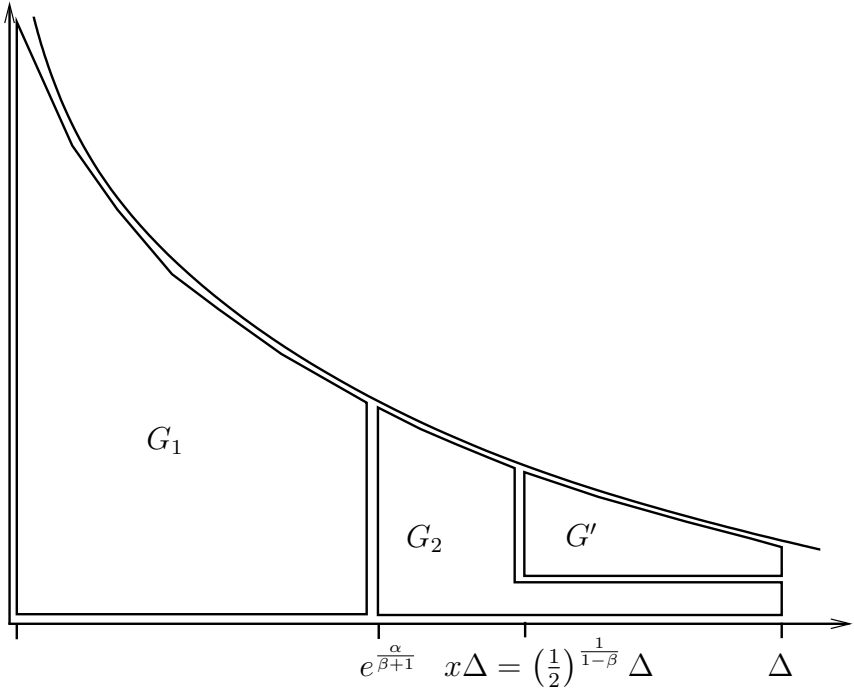
\end{center} 
\caption{Construction of the embedding for the case of $\beta <1$}\label{figure1}
\end{figure}

Let us now give the details of the construction.
First we choose the parameters $\alpha$ and $0<x<1$ such that 
\begin{equation}\label{scl}
n\:\:\leq\:\: x\Delta\:\: \mbox{and} \:\:n\leq |[x\Delta,\Delta ]|.
\end{equation}  
The first inequality in (\ref{scl}) will enable us to implement the node degrees within $[x\Delta, \Delta ]$ 
by replacing edges $\{(v,1),(v,2)\}$ in $G'$ by multi-edges.
The second condition ensures that the interval $[x\Delta,\Delta ]$ is sufficiently large such that $G'$ can be embedded into it.
Then the size of the node degree interval $[x\Delta, \Delta]$ can be estimated as follows: 
\begin{equation}
|[x\Delta, \Delta]|\:\in\:
\left [ \frac{\Delta}{1-\beta}\cdot \left (1-x^{1-\beta}\right )-(x^{-\beta}-1),
        \frac{\Delta}{1-\beta}\cdot \left (1-x^{1-\beta}\right )\right ]
\end{equation}
This means that we have to choose $x$ such that $x\leq \frac{1-x^{1-\beta}}{1-\beta}$, i.e. 
\begin{equation}\label{nn}
(1-\beta )x\: +\: x^{1-\beta}\: -1\:\:\leq\: 0\:\mbox{with}\: x\in (0,1)
\end{equation}
We observe that (\ref{nn}) holds provided we choose $x$ such that
\begin{eqnarray}\label{nnn}
\max\left\{(1-\beta)x,\: x^{1-\beta} \right\}\:\:\leq\:\: \frac{1}{2}
\end{eqnarray}
which is equivalent to
\begin{eqnarray}\label{nnnn}
x\:\leq\:\min\left\{\frac{1}{2(1-\beta)},\left (\frac{1}{2}\right )^{\frac{1}{1-\beta}} \right\}
\end{eqnarray}
We have 
\[\left (\frac{1}{2}\right )^{\frac{1}{1-\beta}}\leq \frac{1}{2(1-\beta)}\:\:\mbox{or equivalently}\:\: 2\geq 2^{1-\beta}(1-\beta)^{1-\beta},\]
and thus if we choose $x=\left (\frac{1}{2}\right )^{\frac{1}{1-\beta}}$, then (\ref{nn}) holds. 
This yields $|[x\Delta,\Delta ]|=(1-o(1))\frac{1}{2}\cdot\frac{\Delta}{1-\beta}$, and furthermore 
$n=x\Delta=(\frac{1}{2})^{\frac{1}{1-\beta}}\cdot\Delta$.
Now we have to construct the residual graph $G_{\alpha,\beta}\setminus G'$.
A straight forward approach is to construct two auxiliary graphs $G_{[1,x\Delta]}$ and $G'_{x\Delta,\Delta]}$, where  
$G_{[1,x\Delta]}=G_d$ for $d$ being the degree sequence of the interval $[1,x\Delta]$ and $G'_{[x\Delta,\Delta]}=G_{d'}$, where the
sequence  $d'$ is constructed as follows:
$d'$ is a degree sequence for the remaining $|[x\Delta,\Delta]|-n$ nodes in $[x\Delta,\Delta]$. It turns out that in this case, the upper bound for the size of an independent set
in $G'_{[x\Delta,\Delta]}=G_{d'}$ would be too large compared to $n$. 
Therefore we construct the residual graph $G_{\alpha,\beta}\setminus G'$ as follows. 
We split the interval $[1,x\Delta ]$ into two parts and 
construct one auxiliary subgraph $G_1$ for the node degree interval $[1,e^{\alpha\slash (\beta +1)})$
and one subgraph $G_2$ for the degree sequence $d$ consisting of the full interval $[e^{\alpha\slash (\beta +1)},x\Delta-1]$
and the degree sequence for the remaining $|[x\Delta,\Delta]|-n$ nodes within the interval 
$[x\Delta,\Delta]$. 
This construction is also shown in Figure \ref{figure1}.
An upper bound for $IS(G_2)$ is given by the bound for the size of a maximum independent
set of the graph $G_{[e^{\alpha\slash (\beta +1)},\Delta]}=:G_3$. We have
\begin{eqnarray*}
IS(G_3) & \leq & \left\lceil \frac{\sum_{i=e^{\alpha\slash (\beta +1)}}^{e^{\alpha\slash\beta}}\left\lfloor\frac{e^{\alpha}}{i^{\beta}}\right\rfloor}{e^{\alpha\slash (\beta +1)}}\right\rceil
                 \:\:\leq\sum_{i=e^{\alpha\slash (\beta +1)}}^{e^{\alpha\slash\beta}}\frac{e^{\alpha\cdot (1-\frac{1}{\beta +1})}}{i^{\beta}}\\
        & \leq & e^{\alpha\cdot\frac{\beta}{\beta +1}}
                 \cdot \left (\int_{e^{\alpha\slash (\beta +1)}}^{e^{\alpha\slash\beta}+1}x^{-\beta}dx\: 
                 +\frac{1}{e^{\alpha}}-\frac{1}{e^{\alpha\cdot\frac{\beta}{\beta +1}}} \right )
         \: \leq \: \frac{e^{\frac{\alpha}{\beta +1}}}{1-\beta}
\end{eqnarray*}
Now we consider the graph $G_1$. We obtain the following bound for the number of nodes of $G_1$:
\begin{eqnarray*}
|[1,x\Delta]| & \leq & \frac{\Delta}{1-\beta}\cdot\left (x^{1-\beta}-\left (\frac{1}{\Delta}\right )^{1-\beta}\right )
              \: =    \: \frac{x^{1-\beta}}{1-\beta}\cdot\Delta\: -\: e^{\alpha}
\end{eqnarray*}  
The size of a maximum independent set in $G_1$ can be bounded as follows. We split the interval $[1,x\Delta]$ into two subintervals $I_{y,1}=\left [1,y\Delta\right )$ 
and $I_{y,2}=[y\Delta,x\Delta]$,
where $y$ has to be chosen appropriately within the interval $(0,x)$. The size of an independent set in the subgraph $G_1[I_{y,1}]$ 
of $G_1$ induced by the node interval $I_{y,1}=\left [1,y\Delta\right )$ is estimated as follows:
\begin{eqnarray*}
IS(G_1[I_{y,1}]) & \leq & \sum_{i=1}^{y\Delta}\left\lceil \frac{\left\lfloor \frac{e^{\alpha}}{i^{\beta}}\right\rfloor }{i}\right\rceil\:
                   \leq\: \sum_{i=1}{y\Delta}\frac{e^{\alpha}}{i^{\beta +1}} \: +y\Delta\\
                 & \leq & \left [\frac{e^{\alpha}}{-\beta x^{\beta}} \right ]_{1}^{y\Delta +1} + e^{\alpha}\left (1-\frac{1}{(y\Delta +1)^{\beta +1}} \right )+y\Delta\\
                 & =    & \frac{e^{\alpha}}{\beta}\left (1-\frac{1}{(y\Delta +1)^{\beta}} \right )
                          + e^{\alpha}\left (1-\frac{1}{(y\Delta +1)^{\beta +1}} \right )+y\Delta
\end{eqnarray*}
Moreover, the size of a maximum independent set in the subgraph $G_1[I_{y,2}]$ induced by the node degree interval $I_{y,2}$ can be 
bounded as follows:
\begin{eqnarray*}
IS(G_1[I_{y,2}]) & \leq & \left\lceil\frac{\sum_{i=y\Delta}^{x\Delta}\left\lfloor\frac{e^{\alpha}}{i^{\beta}}\right\rfloor}{y\Delta}\right\rceil
                 \: \leq \: \sum_{i=y\Delta}^{x\Delta}\frac{e^{\alpha}}{y\cdot i^{\beta}\cdot e^{\alpha\slash\beta}}\:\: +\: 1\\
                 & \leq & \frac{e^{\alpha\cdot (1-1\slash\beta)}}{y}
                          \cdot\left (\int_{y\Delta}^{x\Delta +1}\frac{1}{z^{\beta}}dz\: +\frac{1}{(y\Delta)^{\beta}}-\frac{1}{(x\Delta)^{\beta}}\right )+1\\
                 & \leq & \frac{e^{\alpha\cdot (1-1\slash\beta)}}{y}
                          \cdot\left (\frac{(x\Delta)^{1-\beta}}{1-\beta}-\frac{(y\Delta)^{1-\beta}}{1-\beta}  +\frac{1}{(y\Delta)^{\beta}}-\frac{1}{(x\Delta)^{\beta}}\right )+1\\
                 & =    & (1\pm o(1))\cdot\frac{e^{\alpha}}{y\Delta}\cdot \frac{\Delta^{1-\beta}}{1-\beta}\left (x^{1-\beta}-y^{1-\beta}\right )\:\: =\:\: \Theta (y^{-1})
\end{eqnarray*}
Thus we obtain the following estimate for the size of a maximum independent set in the graph $G_1$: 
\begin{equation}\label{bbbb}
IS(G_1)\:\: =\:\: \Theta \left (y\cdot\Delta\: +\: \frac{1}{y}\right )
\end{equation}
In order to obtain an upper bound for the right hand side in (\ref{bbbb}), we observe that 
\[\frac{d}{dy}\left (\frac{1}{y}+y\cdot\Delta\right )\: =\: \Delta-\frac{1}{y^2},\]
Thus we choose $y=\Theta \left (\frac{1}{\sqrt{\Delta}}\right )$, say $y=\Delta^{-1\slash 2}$.
This yields $IS(G_1)=\Theta (\sqrt{\Delta})$. 

Now we start from a graph $G\in {\mathcal G}_{\frac{1}{n^{1-\epsilon}}}\cup {\mathcal G}_{\frac{1}{n^{\epsilon}}}$.
We construct the graph $G'$. Let $n$ be the number of nodes of $G'$. Thus we have 
\begin{equation}
G'\in {\mathcal G}_{\frac{2^{\epsilon}}{n^{1-\epsilon}}}\cup {\mathcal G}_{\frac{1}{2^{1-\epsilon}n^{\epsilon}}}
\end{equation}
Then we choose the parameters $x$ and $\alpha$:
\begin{equation}
x=\left (\frac{1}{2}\right )^{\frac{1}{1-\beta}},\quad\: \alpha=\beta\cdot \ln\left (\frac{n}{x}\right )
\end{equation}
We construct the graph $G_{\alpha,\beta }$ as described above.
In particular, $G'$ will be embedded into the node degree interval $[x\Delta, \Delta ]$, which is of size
at least $(1-o(1))\cdot\frac{\Delta}{1-\beta}(1-x^{1-\beta})$. 
 
Altogether we obtain the following result. 
\begin{theorem}
For every $\beta\in (0,1)$, for every $\epsilon >0$ the Maximum Independent Set Problem on $(\alpha,\beta )$-PLGs is NP-hard to approximate within $n^{1-\epsilon}$.
\end{theorem} 
\section{The Case $\beta =1$}
Now we consider the case when the power law exponent is equal to $1$. First, we observe that in this case, there is a simple $O(\ln (n))$-approximation algorithm 
for the Independent Set Problem in $(\alpha,1)$-PLGs. Namely, the number of nodes is $\alpha e^{\alpha}$, and we have $e^{\alpha}$ nodes of degree $1$. Thus taking half of them 
yields a $O(\ln (n))$-approximate independent set.\\[1ex]
We will now give a lower bound of the form $\Delta^{\epsilon}$ for maximum degree $\Delta$ and some constant $\epsilon >0$
for the Independent Set Problem in $(\alpha,1)$-PLGs. 
As in the previous case $\beta <1$, we want to proceed as follows: We start from a class ${\mathcal G}$ of graphs for which the Independent Set Problem is hard to approximate.
Then we construct a polynomial time reduction which embeds every $G\in {\mathcal G}$ into an $(\alpha,1)$-PLG $G_{\alpha,1}$.
Since the upper bound is $O(\log n)$, we should choose ${\mathcal G}$ appropriately such that the approximation lower bound for Independent Set restricted to graphs 
from ${\mathcal G}$ is at most logarithmic.  

In \cite{alon95} it was shown that the Independent Set Problem on graphs of degree bounded by $\Delta$ is hard to approximate within $\Delta^{\epsilon}$ for some fixed
$\epsilon >0$. This result also holds for $\Delta =\Theta (\log n)$. We will embed those graphs into $(\alpha,1)$-PLGs $G_{\alpha,1}$. The construction in \cite{alon95}
starts from a class of graphs ${\mathcal G}={\mathcal G}_a\cup {\mathcal G}_b$, where $\alpha (G)\leq a\cdot n$ for $G\in {\mathcal G}_a$ and $\alpha (G)\geq b\cdot n$ for all
$G\in {\mathcal G}_b$ and $0<a<b<1$ are constant. Given a graph $G\in {\mathcal G}$ with $n$ vertices, a degree $d$-bounded Ramanujan graph $H$ on $n$ vertices is chosen, and 
the graph $\tilde{DG}^k$ is constructed as follows. The vertices of $\tilde{DG}^k$ are walks $(v_1,\ldots , v_k)$ of length $k$ in $H$. Two such vertices 
$v=(v_1,\ldots , v_k)$ and $u=(u_1,\ldots , u_k)$ are connected by an edge iff the vertex subset $\{v_1,\ldots , v_k,u_1,\ldots , u_k\}$ is not an independent set in $G$.
In \cite{alon95}, Theorem 2.1 it is shown that $\alpha (\tilde{DG}^k)$ is within the interval 
\[\begin{array}{l}
\left [\:\alpha (G)d^{k-1}\left (\frac{\alpha (G)}{n}+\lambda_{n-1}\left (1-\frac{\alpha (G)}{n}\right )\right )^{k-1},\:
\alpha (G)d^{k-1}\left (\frac{\alpha (G)}{n}+\lambda_{1}\left (1-\frac{\alpha (G)}{n}\right )\right )^{k-1}\right ]
\end{array}\]
where $A$ is the transition matrix of the random walk on the Ramanujan graph $H$, $\lambda_0\geq\ldots\geq\lambda_{n-1}$ are the eigenvalues of $A$ (note that $A$ is 
symmetric and has only real eigenvalues) and we have $\lambda_0=1,\lambda:=\max\{\lambda_1,|\lambda_{n-1}|\}\leq 2\sqrt{d-1}\slash d$.

\begin{figure}[htb]
\begin{center}
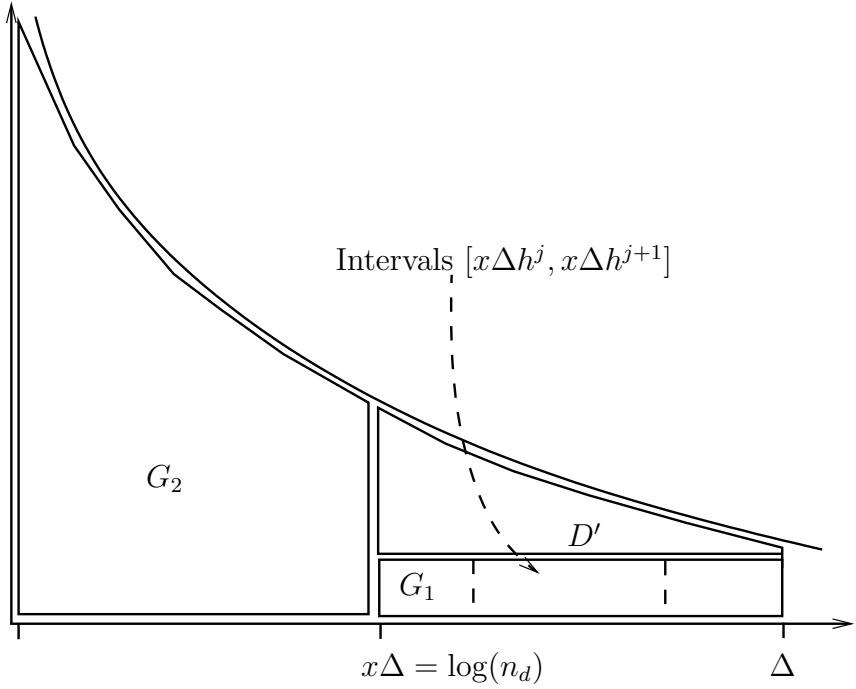
\caption{Construction for the case $\beta=1$}\label{figure2}
\end{center}
\end{figure}

Now we want to choose $k$ appropriately such that the maximum degree of $\tilde{DG}^k$ is logarithmic in the number of its nodes.
The number of nodes of $\tilde{DG}^k$ is $n\cdot d^{k-1}$. The maximum degree is $d^{k-1}\cdot 3k^2$. Thus we have to choose $k$ such that for some constant $c>0$,
\[\begin{array}{c@{\:}r@{\:}c@{\:}l}
                & d^{k-1}\cdot 3k^2 & \leq & c\cdot \log\left (n\cdot d^{k-1}\right )\\
\Leftrightarrow & d^{k-1}\cdot 3k^2 & \leq & \log (n)+(k-1)\log (d)+\log (c)\\
\Leftrightarrow & (k-1)\log (d)+\log (3k^2) & \leq & \log\log (n) +\log\left (1+\frac{(k-1)\log (d)+\log (c)}{\log (n)}\right )
\end{array}\]
We observe that for $k_l=\frac{\log\log (n)}{3\ln (d)}$ we obtain 
\begin{equation}
\Delta_l=d^{k-1}\cdot 3k^2\approx (\log n)^{1\slash 3}\cdot\frac{(\log\log n)^2}{(\ln d)^2}
\end{equation}
For $k_u=\frac{\log\log (n)}{\ln d}$ we obtain
\begin{equation}
\Delta_u=d^{k-1}\cdot 3k^2\approx e^{\log\log (n)}\cdot 3\cdot \frac{(\log\log (n))^2}{(\ln d)^2} = \log (n)\cdot\frac{3(\log\log n)^2}{(\ln d)^2}
\end{equation}
Thus we choose $k\in [k_l,k_u]=\left [\frac{\log\log n}{3\ln d},\frac{\log\log n}{\ln d} \right ]$ appropriately such that $\Delta_k=d^{k-1}\cdot 3k^2$
satisfies $\Delta_k=\log n$. We let $D$ denote the product graph $\tilde{DG}^k$. 
Now we have to choose parameters $\alpha$ and $x$ such as to embed the graph $D$ into the node degree interval $[x\Delta,\Delta ]$, where $\Delta=e^{\alpha}$
is the maximum degree of the $(\alpha, 1)$-PLG $G_{\alpha, 1}$. Let $n_d=n\cdot d^{k-1}$ be the number of nodes of $D$.
Thus we have to meet the following conditions:
\begin{itemize}
\item[(I)] $n_d\:\leq\: |[x\Delta,\Delta]|\:\approx\: e^{\alpha}\ln\left (\frac{1}{x}\right )$
\item[(II)] $\log (n_d)\:\leq\: x\Delta$
\end{itemize}
Thus we choose $x=\frac{\log (n_d)}{e^{\alpha}}$. 
In order to satisfy the requirement (II), we 
choose $\alpha =\log (n_d)$. Then (I) holds as well.
We obtain 
\begin{equation}
|[x\Delta,\Delta]|\approx e^{\alpha}\ln\left (\frac{1}{x} \right ) = e^{\alpha}\cdot \alpha-e^{\alpha}\cdot \ln (\alpha )=(1-o(1))\alpha e^{\alpha}
\end{equation}
Now we have to give an estimate for the size of a maximum independent set in the subgraph $G_1$ induced by the residual set of nodes $[x\Delta,\Delta ]\setminus V(G)$.
This can be upper-bounded by $IS([x\Delta, \Delta])$. Thus it remains to give an estimate for $IS([x\Delta, \Delta])$, the size of a maximum independent
set of a graph implementing the complete interval $[x\Delta, \Delta]$ of the power law distribution with parameters $\alpha$ and $1$. If we would use the inequality
\begin{equation}
IS([x\Delta,\Delta ])\:\:\leq\:\:\left\lceil \frac{\sum_{i=x\Delta}^{\Delta}\left\lfloor\frac{e^{\alpha}}{i}\right\rfloor}{x\Delta}\right\rceil,
\end{equation}
the resulting bound would be $e^{\alpha}$ plus some lower order terms, which is not sufficient for our purpose.
We proceed as follows: We split the interval $[x\Delta,\Delta ]$ into $L$ subintervals of the form $[x\Delta\cdot h^j,x\Delta\cdot h^{j+1}], j=0,\ldots L-1$ and use the estimate
\begin{equation}\label{betaeinseq}
IS([x\Delta, \Delta ])\:\leq\:\sum_{j=0}^{L-1}\left\lceil\frac{\sum\limits_{i=x\Delta\cdot h^j}^{x\Delta\cdot h^{j+1}}
\left\lfloor \frac{e^{\alpha}}{i}\right\rfloor }{x\Delta\cdot h^j}\right\rceil
\end{equation}
where $h$ and $L$ are parameters of this estimate.
This means that we have to choose $h^L=\frac{1}{x}=\frac{e^{\alpha}}{\alpha}$, i.e. $h=\sqrt{e^{\alpha}\slash\alpha}$.
The whole construction for the case $\beta =1$ is shown in Figure \ref{figure2}.
%
%
Now the right hand side in (\ref{betaeinseq}) is
\begin{eqnarray*}
 & \leq & \sum_{j=0}^{L-1}\sum_{i=x\Delta\cdot h^j}^{x\Delta\cdot h^{j+1}}\frac{e^{\alpha}}{i\cdot x\Delta\cdot h^j}\:\:\: +L
  =     \sum_{j=0}^{L-1}\sum_{i=x\Delta\cdot h^j}^{x\Delta\cdot h^{j+1}}\frac{1}{x\cdot h^j\cdot i}\:\:\: +L
\end{eqnarray*}
We approximate the inner sum by an integral. This yield that the right hand side in (\ref{betaeinseq}) is bounded by
\begin{eqnarray*}
 &  & \sum_{j=0}^{L-1}\frac{1}{xh^j}\cdot\left (\int_{x\Delta h^j}^{x\Delta h^{j+1}}\frac{1}{y}dy\:\: +\frac{1}{x\Delta h^{j+1}}-\frac{1}{x\Delta h^j}\right ) 
          \:\: +L\\
 & =    & \sum_{j=0}^{L-1}(1+o(1))\cdot \frac{1}{xh^j}\cdot\left (\ln (x\Delta h^{j+1})-\ln (x\Delta h^j)+\frac{1}{x\Delta h^{j+1}}-\frac{1}{x\Delta h^j}\right )
          \:\: +L\\
 & =    & \sum_{j=0}^{L-1}(1+o(1))\cdot \frac{e^{\alpha (1-\frac{j}{L})}}{\alpha^{1-\frac{j}{L}}}\cdot 
          \left (\frac{j+1}{L}\cdot\alpha-\frac{j}{L}\cdot\alpha \right )\\
 & =    & \sum_{j=0}^{L-1}(1+o(1))\frac{e^{\alpha (1-\frac{j}{L})}\cdot\alpha^{\frac{j}{L}}}{L}\: 
          =\: (1+o(1))\frac{e^{\alpha}}{L}\sum_{j=0}^{L-1}\left (\frac{\alpha}{e^{\alpha}}\right )^{\frac{j}{L}}\\
 & =    & (1+o(1))\frac{e^{\alpha}}{L}\cdot\frac{1-\left (\frac{\alpha}{e^{\alpha}} \right )^{\frac{L}{L}}}{1-\left (\frac{\alpha}{e^{\alpha}} \right )^{\frac{1}{L}}}
          =(1+o(1))\frac{e^{\alpha}}{L}\cdot\frac{1-\frac{\alpha}{e^{\alpha}}}{1-\left (\frac{\alpha}{e^{\alpha}} \right )^{\frac{1}{L}}}
\end{eqnarray*} 

Now we choose $L=\alpha$. First we observe that 
\begin{equation}
\log\left ((\frac{\alpha}{e^{\alpha}} )^{1\slash\alpha}\right )=\frac{1}{\alpha}(\log (\alpha )-\alpha)=\frac{\log\alpha}{\alpha}-1\:\:\longrightarrow\:\: -1\:\: (\alpha\to\infty)
\end{equation}
and therefore $\left (\frac{\alpha}{e^{\alpha}}\right )^{\frac{1}{\alpha}}$ converges to some constant $c'\in (0,1)$, as $\alpha$ tends to infinity. 
Thus we obtain $|IS([x\Delta,\Delta ])|\leq c\cdot\frac{e^{\alpha}}{\alpha}$ for some constant $c$.

Now we give an upper bound for the size of an independent set within $[1,x\Delta]=[1,\alpha]$:  
\begin{eqnarray*}
IS([1,x\Delta]) & \leq & \sum_{i=1}^{\alpha}\left\lceil\frac{\left\lfloor\frac{e^{\alpha}}{i}\right\rfloor}{i}\right\rceil\:\leq\: \sum_{i=1}^{\alpha}\frac{e^{\alpha}}{i^2}\: +\alpha\\
                & \leq & \alpha\: +\: e^{\alpha}\left (\int_{1}^{\alpha}\frac{1}{y^2}dy\: +\: 1-\frac{1}{\alpha^2} \right )\\
                & \leq & \alpha\: +\: e^{\alpha}\left ( 1-\frac{1}{\alpha}+1-\frac{1}{\alpha^2}\right )\:\: =\:\: (1-o(1))\cdot 2e^{\alpha}
\end{eqnarray*}
Now we start from a graph $G\in {\mathcal G}_a\cup {\mathcal G}_b$ with $n$ nodes. We construct the graph $D=\tilde{DG}^k$ for $k=\log\log n$.
Thus the maximum degree of $D$ is $\Delta_k=d^{k-1}\cdot 3k^2$, where $d$ is the degree of the Ramanujan graph used in the construction of 
$D$. The resulting lower bound for the approximability of Maximum Independent Set in graphs of the form $D=\tilde{DG}^k,G\in {\mathcal G}_a\cup {\mathcal G}_b$ is
\[bnd^{k-1}\cdot\left (b+\frac{2\sqrt{d-1}}{d}\right )^{k-1}\slash\left (and^{k-1}\cdot\left (a-\frac{2\sqrt{d-1}}{d}\right )^{k-1} \right )
=\frac{b}{a}\cdot\left (\frac{b-\epsilon_2}{a+\epsilon_2}\right )^{k-1}\]
where we can choose $\epsilon_2$ arbitrary small by choosing $d$ sufficiently large. 
Now $D$ is embedded into an $(\alpha, 1)$-PLG $G_{\alpha, 1}$ (cf. Figure \ref{figure2}). The number of nodes of $G_{\alpha, 1}$ is equal to $\alpha e^{\alpha}$,
and $G_{\alpha, 1}$ consists of the subgraphs $D'$, $G_1$ and $G_2$. We have shown that the size of a maximum independent set in the subgraph $G_1$ is of order 
$\frac{1}{\alpha}e^{\alpha}$, while the size of a maximum independent set in the subgraph $G_2$ is of order $e^{\alpha}$. Thus it remains to show that 
if $G\in {\mathcal G}_{b}$, then this implies that $IS(D')\geq \alpha^{\epsilon}\cdot e^{\alpha }$ for some constant $\epsilon$.
It suffices to show that
\begin{equation}\label{ffff}
(b+\epsilon_2)^{k-1}\:\: >\:\: \frac{1}{(\log (nd^{k-1}))^{1\slash\epsilon}}
\end{equation} 
We let $B'=\frac{1}{b+\epsilon_2}$. Now (\ref{ffff}) is equivalent to 
\begin{equation}\label{fffff}
(\log n)^{1\slash\epsilon }\left (1+\frac{(k-1)\log d}{\log (n)} \right )^{1\slash\epsilon }\: >\: B'^{k-1},
\end{equation}
and (by taking logarithms) to $\frac{1}{\epsilon}\cdot\log\log (n)\cdot (1+o(1))> (k-1)\cdot\log (B')$. 
We observe that for $k=\log\log n$ there exists some $\epsilon >0$ such that this last inequality holds.  
Thus we obtain the following result.
\begin{theorem}
For $\beta =1$, there exists an $\epsilon \in (0,1)$ such that the Maximum Independent Set Problem on $(\alpha,\beta )$-PLGs is NP-hard to approximate within $(\log n)^{\epsilon}$.
\end{theorem}
{\bf Remark:} In this section we have constructed an embedding of graphs $G\in {\mathcal G}={\mathcal G}_a\cup {\mathcal G}_b$ into $(\alpha,1)$-PLGs, for $0<a<b<1$ being constant.
It combines the construction from \cite{alon95} with an embedding of the resulting MIS instances $D=\tilde{DG}^k$ into power law graphs for $\beta =1$. 
This construction does not work for the case of $a,b$ being non-constant, e.g. if we try to start from the class of graphs 
${\mathcal G}={\mathcal G}_{1\slash n^{1-\epsilon}}\cup {\mathcal G}_{1\slash n^{\epsilon}}$ from \cite{zuckerman06}. The reason is that in the construction from \cite{alon95},
a $d$-regular Ramanujan graph $H$ is used, where $d>\frac{16}{(b-a)^2}$. 
For $a=1\slash n^{1-\epsilon},b=1\slash n^{\epsilon}$ this yields a non-constant
lower bound for $d$, which implies that the graph $D$ would be super-polynomially larger than $G$.   
\section{Further Research}
We have given new approximation lower bounds for the MIS problem in $(\alpha,\beta )$-PLGs for $\beta\leq 1$. For $\beta <1$ being constant, the
lower bound is $n^{1-\epsilon}$ for every $\epsilon$, while for $\beta =1$ the lower bound is $(\log n)^{\epsilon}$ for some
constant $\epsilon\in (0,1)$. The further improvements on these lower bounds are important open questions in this area. Another question is 
the status of the functional cases around value $\beta =1$, i.e. when $\beta$ is of the form $\beta_f=1\pm\frac{1}{f(n)}$
for some function $f(n)$ depending on the size $n$ of the graph. Another question is the approximability status of the MIS on random PLGs
in the preferential attachment model \cite{barabasi}.


\end{document}